\documentclass[aps,prb,twocolumn,10pt]{revtex4-1}
\usepackage{longtable}
\usepackage{subfigure}
\usepackage{amsmath,amssymb,amsfonts,bm}
\usepackage{graphicx}
\usepackage{ulem}
\usepackage{color}

\def\KeyWord#1{$\backslash$\IfColor{$\!\!$\textRed{#1}\textBlack}{#1}$\!\!$}
\newcommand{\be}{\begin{equation} }
\newcommand{\ee}{\end{equation} }
\newcommand{\ba}{\begin{eqnarray} }
\newcommand{\ea}{\end{eqnarray} }

\newcommand{\bit}{\begin{itemize}}
\newcommand{\eit}{\end{itemize}}

\newcommand{\ve}{\varepsilon }

\def\em{\it}

\newcommand{\oS}{ \mathbf{\hat{S}}}

\graphicspath{{../Figures/}}

\begin{document}

\title{Exactly Soluble Model of a 3D Symmetry Protected Topological Phase of Bosons with Surface Topological Order}

\author{F. J. Burnell$^{1,2}$}
\author{ Xie Chen$^{3,4}$}
\author{ Lukasz Fidkowski$^{3,5}$}
\author{Ashvin Vishwanath$^{3}$}
\affiliation{1. Department of Physics and Astronomy, University of Minnesota, Minneapolis, MN, 55455, USA}
\affiliation{2. Rudolf Peierls Centre for Theoretical Physics, 1 Keble Road, Oxford, OX1 3NP, United Kingdom}
\affiliation{3. Department of Physics, University of California, Berkeley, California 94720, USA}
\affiliation{4.  Department of Physics, California Institute of Technology, Pasadena, CA 91125, USA}
\affiliation{5.  Department of Physics and Astronomy, Stony Brook University, Stony Brook, NY 11794-3800, USA.}

\begin{abstract}
We construct an exactly soluble Hamiltonian on the D=3 cubic lattice, whose ground state is a topological phase of bosons protected by time reversal symmetry, i.e a symmetry protected topological (SPT) phase.  In this model, excitations with  anyonic statistics are shown to exist at the surface  but not in the bulk. The statistics of these surface anyons is explicitly computed and shown to be identical to the 3-fermion $\mathbb{Z}_2$ model, a variant of $\mathbb{Z}_2$ topological order which cannot be realized in a purely D=2 system with time reversal symmetry. Thus the model realizes a novel surface termination for 3D SPT phases, that of a fully symmetric gapped  surface with topological order. The 3D phase found here was previously proposed from a field theoretic analysis but is outside the group cohomology classification that appears to capture all SPT phases in lower dimensions. Such phases may potentially be realized in spin-orbit coupled magnetic insulators, which evade magnetic ordering. 
Our construction utilizes  the Walker-Wang prescription to create a 3D confined phase with surface anyons, which can be extended to other topological phases. 
\end{abstract}
\maketitle

\section{Introduction}
Recently, there has been much progress in understanding topological phases of interacting bosons~\cite{Pollmann2012,Chen2011a,Schuch2011,Turner2011,Fidkowski2011,Chen2011,Chen2012a,Levin2012,Lu2012a,Vishwanath2012} that are short ranged entangled (SRE) - i.e. which have a gapped bulk that is free of exotic excitations\cite{Note2} - but which are nevertheless distinct from the trivial phase in the presence of a symmetry.  Such symmetry protected topological phases (SPTs) are significantly simpler than intrinsically topologically ordered phases, such as fractional quantum Hall states and gapped spin liquids, whose bulk anyonic excitations reflect their long range entangled nature\cite{Wen2004B, Note1}. They naturally generalize the notion of free fermion topological insulators and superconductors  \cite{Hasan2010,Qi2011,Hasan2011} to interacting bosonic systems. An experimental example is the Haldane S=1 antiferromagnet in 1+1 dimension, protected by spin rotation symmetry \cite{Haldane1983b,Affleck1987}

In a recent breakthrough, analogous states were shown to exist in higher dimensions\cite{Chen2011,Chen2012a,Levin2012,Lu2012a,Vishwanath2012}, which could potentially be realized as ground states of frustrated magnetic insulators or ultra cold bosonic atoms\cite{Senthil2012, Lu2012c,Ye2012,Geraedts2012}. The simplest example is a 2+1D bosonic phase with a gapped bulk but  $c_-=8 n $ ($n$ an integer) edge modes that all propagate in the same direction \cite{Kitaev2006,Note2}.  We will refer to the $n=1$ member of this sequence as the Kitaev $E_8$ state - which is a bosonic analogue of the fermionic $p_x+ip_y$ superconductor.  With symmetries, more phases are possible, and it was proposed \cite{Chen2011,Chen2012a} that they are classified by a fundamental mathematical object associated with the symmetry group $G$ -- namely the cohomology groups $H^{d+1}(G,U(1))$ in $d+1$ dimensions\cite{Chen2011,Chen2012a}.  While this assertion was verified in several cases by other means \cite{Levin2012,Lu2012a,Levin2012a,Vishwanath2012,Xu2012}, intriguingly, the field theoretical approach\cite{Vishwanath2012} predicted an additional SPT phase in 3+1 D protected by time reversal symmetry $\mathcal T$. This state is a bosonic analogue of the 3+1D free fermion topological superconductor (class DIII)
hence referred to as the 3D BTSc. 
While the BTSc was discussed as a physical possibility in Ref. \cite{Vishwanath2012}, that work did not definitively establish it as a phase of matter.  Specifically, the chiral nature of the field theory in question lead to questions about whether the BTSc could in fact be realized on the lattice (a worry that does not apply to other similar phases discussed in Ref. \cite{Vishwanath2012}).  Here we irrefutably demonstrate its existence via a realization in an exactly soluble model, which also naturally exhibits an exotic surface state. 

Further, as opposed to the edge of a 2+1D SPT, which must either be gapless or spontaneously break symmetry, the 2D surface of a 3+1D SPT allows for a novel possibility: a fully gapped and symmetric state is allowed if the surface develops topological order\cite{Vishwanath2012}.  However, this surface state is anomalous - i.e. it implements the global symmetries in a way that cannot be realized in a strictly 2+1D phase.  In the context of the 3D BTSc, the topologically ordered surface is conjectured\cite{Vishwanath2012} to realize the ``3-fermion $\mathbb{Z}_2$ state",  {\it with} $\mathcal T$.  This is a variation on ${\mathbb Z}_2$ gauge theory in which all three particles - $e$,$m$,$\ve$ - are fermions with mutual $\pi$ statistics.  A strictly 2D realization of this 3-fermion $\mathbb{Z}_2$ state always breaks $\mathcal T$ since it is associated with $c_-=4 \,\rm{mod}\,8$ chiral edge boson modes\cite{Kitaev2006}.  Our exactly soluble $3+1$D model proves that the conjecture of Ref. \cite{Vishwanath2012} is correct: its surface harbors precisely this $3$-fermion state while preserving $\mathcal T$.  The key idea is that only the self and mutual statistics of the anyons go into defining the exactly soluble model, and these are all real ($\pm 1$), allowing the $3+1$D model to be $\mathcal T$ invariant.  If $\mathcal T$ is broken on the surface, then domain walls acquire $c_-=8$ chiral edge modes, identical to the edge of the Kitaev E$_8$ state. 

Our model is a special case of the Walker-Wang\cite{Walker2012,Keyserlingk2013} prescription, applied for the first time to produce an SPT phase.  However, rather than using this formalism, we begin by defining our Hamiltonian at an elementary level and explicitly demonstrating its bulk and edge properties. Our Hamiltonian is a spin model  with specially tuned interactions to allow for exact solvability.  Time reversal is the only symmetry considered, which we show remains unbroken in the ground state. Hence this phase models a topological paramagnet (in an insulating system with magnetic moments), in which the spin symmetry is broken down to just time reversal, as would be expected with strong spin-orbit couplings. In contrast to quantum spin liquids, which have exotic deconfined excitations in the bulk, here the unusual physics appears only at the surface. An important future direction is the construction of more realistic magnetic Hamiltonians that lead to this topological paramagnet phase.

\section{ Exactly Soluble Model of a 3D SPT Phase with Surface Topological Order}
Our model is built out of $4$-dimensional spin Hilbert spaces living on the links of a cubic lattice.  We use the following ordered basis for the spin Hilbert space: $\{|1\rangle, |e\rangle,|m\rangle,|\ve\rangle\}$, but also find it convenient to express it as the product of two spin $1/2$'s, acted on by Pauli matrices $\sigma^j$ and $\tau^j$.  In this notation, the ordered basis becomes $\{|++\rangle, |-+\rangle, |+-\rangle, |--\rangle \}$, where the first (second) sign corresponds to the eigenvalue of the Pauli matrix $\sigma^x$ ($\tau^x$).  We will label the particular link with a subscript where necessary.  The Hamiltonian is a sum of vertex ($A_V$) and plaquette ($B_P$) terms:
\be \label{Ham}
H = - \sum_V A_V - \sum_P B_P
\ee
where the first sum is over all vertices $V$ and the second sum is over all plaquettes $P$.  The vertex term is defined as
\be
A_V=  \prod_{i \in *V}\sigma^x_i + \prod_{i \in *V} \tau^x_i
\ee
where $*V$ is the set of $6$ links adjacent to the vertex $V$.

\begin{figure}[htp]
\includegraphics[width=0.6\linewidth]{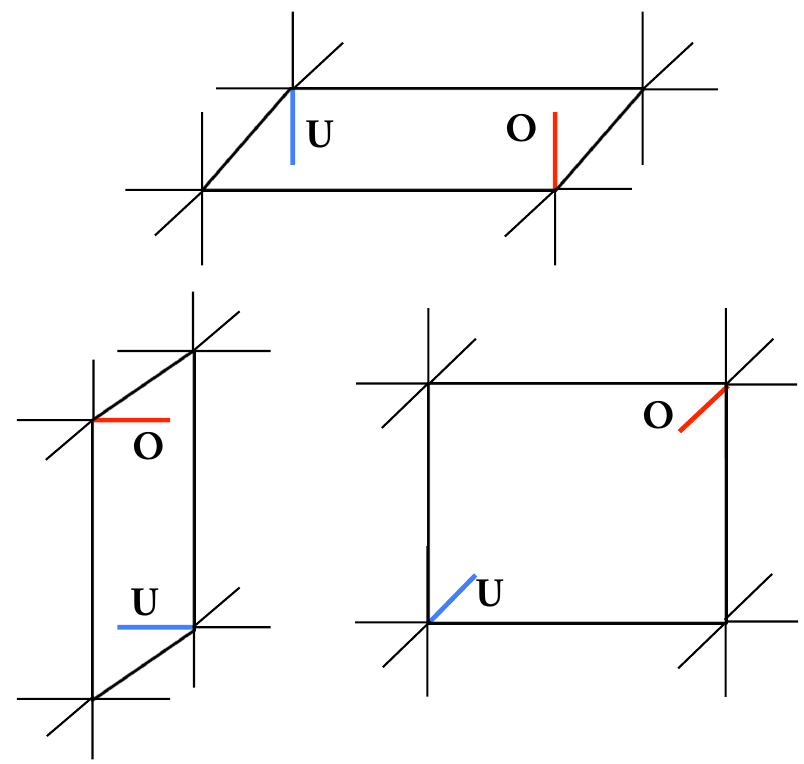}
 \caption{ Choice of links on which $B_P^{(e/m)}$ act, for the three different types of plaquettes in the lattice.  In the chosen projection O links (red) cross over the plaquette $P$,  and U links (blue) cross under it.}
\label{BpFig}
 \end{figure}

The plaquette term is more complicated.  To define it, we fix a specific 2d projection of our 3d lattice once and for all, one that in particular has the property that each plaquette has one of the three forms shown in Fig. \ref{BpFig}.  For each such plaquette $P$, there are two links which end up in its interior under the 2D projection.  These links, labeled $O$ and $U$ in the figure, lie ``over" and ``under" $P$, respectively \cite{Note5}.  The plaquette term $B_P$ then acts on the four links that make up $P$ (we will denote this collection of 4 links by $\partial B$), but also depends on the labels of the associated $O$ and $U$ links.  Specifically, $B_P = B_P^{(e)} + B_P^{(m)}$ where
\begin{eqnarray} \label{eq:bpe}
B_P^{(e)}=\sigma_{O}^x \sigma_{U}^x \tau_{U}^x \prod_{ i \in \partial P } \sigma^z_i
\end{eqnarray}
and
\begin{eqnarray} \label{eq:bpm}
B_P^{(m)}=\sigma_{O}^x \tau_{O}^x \tau_{U}^x \prod_{ i \in \partial P } \tau^z_i.
\end{eqnarray}
To gain some intuition for this Hamiltonian, we can view it as a ``twisted" product of two ${\mathbb Z}_2$ gauge theories.  Indeed, $\sigma_i^x$ and $\tau_i^x$ define a two independent ${\mathbb Z}_2$ charges on each link, which we denote ${\mathbb Z}_2^{(e)}$ and ${\mathbb Z}_2^{(m)}$ respectively.  The vertex terms $A_V$ then simply enforce conservation of ${\mathbb Z}_2^{(e)} \times {\mathbb Z}_2^{(m)}$ charge at each vertex, whereas $B_P$ is the usual ${\mathbb Z}_2$ gauge theory plaquette term twisted by some signs related to the occupation numbers of the $O$ and $U$ links.

An important point is that all of the terms in the Hamiltonian commute.  Indeed, all the vertex terms $A_V$ clearly commute with each other, and since each $B_P$ can only change the $\mathbb{Z}_2^{(e/m)}$ charge on an even number of links adjacent to each vertex (namely $0$ or $2$), the plaquette terms also commute with all the vertex terms.  
To see that $\left[ B_{P_1}, B_{P_2} \right ] =0$, we note that this is clearly true if the O and U links of $P_1$ have no overlap with $\partial P_2$ (note that this is equivalent to the condition with $1$ and $2$ exchanged).  When this condition fails, it must be that either the $O$ link of $P_1$ intersects $\partial P_2$ and the $U$ link of $P_2$ intersects $\partial P_1$, or we have this situation with $1$ and $2$ exchanged.  In both cases, the minus signs from commutators of $x$ and $z$ Pauli matrices cancel in pairs, so $B_{P_1}$ and $B_{P_2}$ commute.

Also, since the matrix elements of $H$ are real, the Hamiltonian is invariant under time reversal $\mathcal T$, where $\mathcal T$ is defined to be complex conjugation of the many body wave function in our $1,e,m,\ve$ basis.  Note that this time reversal operator satisfies $\mathcal T^2=1$.

\subsection{ Trivial Bulk:}  
\begin{figure}[htp]
\includegraphics[width=0.8\linewidth]{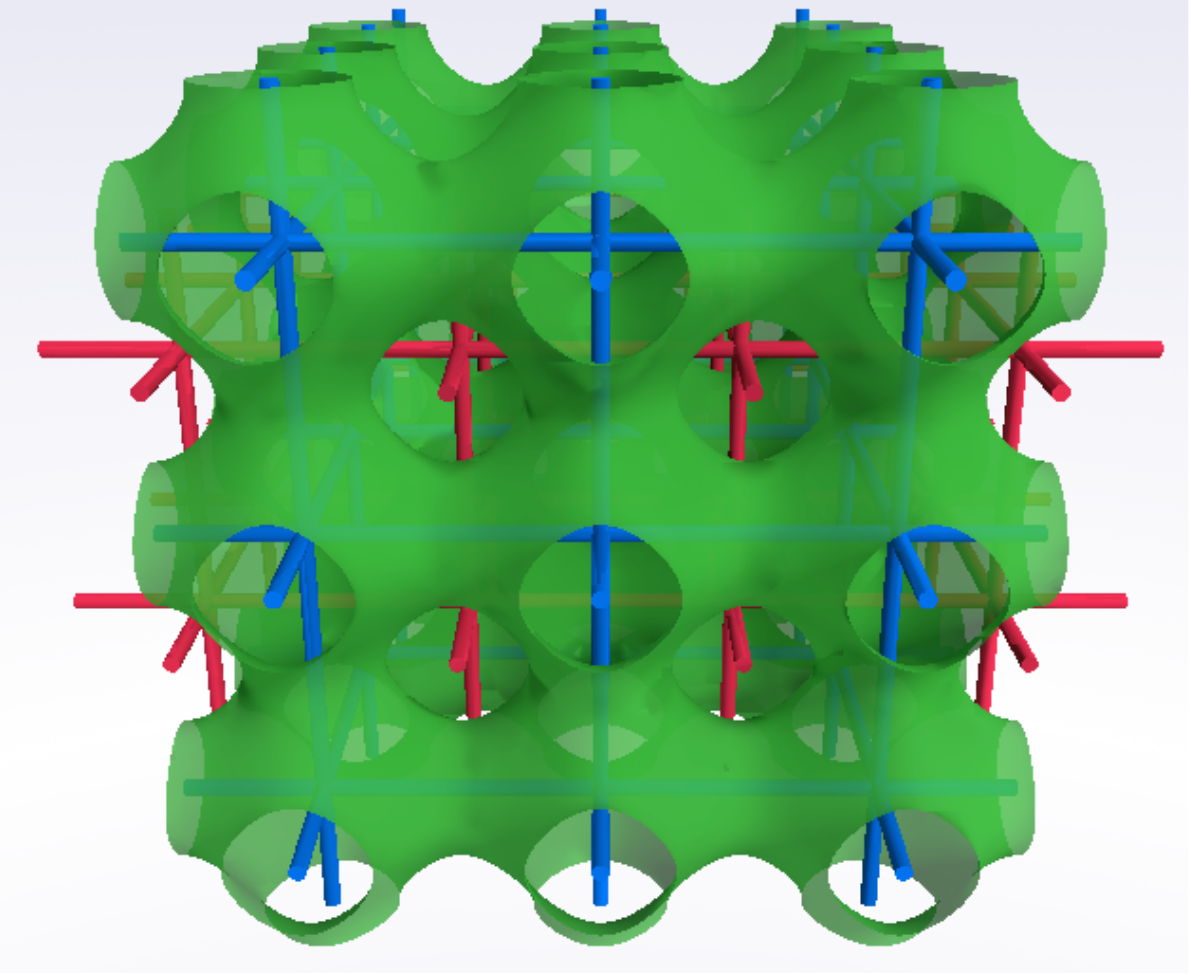}
 \caption{ The plumber's nightmare geometry.  Our model is defined on the links of the blue lattice; its dual lattice is displayed in red.}
\label{pnightmare}
 \end{figure}

We now argue that our model has a unique ground state when defined on topologically non-trivial manifolds.  For definiteness we will work with a 3D torus $T^3$, but our argument generalizes to any orientable 3D manifold.  The first step is to introduce an auxilliary geometry -- the ``plumber's nightmare" shown in Fig. \ref{pnightmare} --a genus $N$ surface which is topologically just the surface of a thickened version of the cubic lattice on which our model is defined (see also \cite{Gils}).  We consider a 2D gapped chiral state with the three fermion topological order defined on this $2$-manifold $M$ (this 2D state does break $\cal T$).  
 The low energy description of this theory is just a $U(1)$ Chern-Simons theory, with $K$-matrix equal to the Cartan matrix of $SO(8)$:
 \begin{eqnarray}
4\pi S_{\rm TQFT} &=& \int d^3 x \sum_{I,J=1}^4 K^{SO(8)}_{IJ} \epsilon^{\mu\nu\lambda} a^I_\mu \partial_\nu a^J_\lambda \\
K^{SO(8)} &=& \left (  \begin{array}{cccc} 2 &-1 &-1 & -1 \\ -1 & 2 & 0 & 0\\ -1 & 0 & 2 & 0 \\ -1 & 0& 0& 2 \end{array}\right )
\end{eqnarray}
 (See Appendix \ref{AppA} for more details).   A key observation is that the low energy sector of this 2D chiral theory defined on $M$ -- which is just a high genus orientable $2D$-manifold -- has dimension exponentially large in $N$, and in fact maps exactly onto the subspace ${\cal H}_v \subset {\cal H}$ with all vertex terms $A_V$ imposed.  Indeed, the ${1,e,m,\ve}$ label on any link just represents the topological charge flowing through the tube enclosing this link, and $A_V$ enforces conservation of topological charge at vertex $V$.

\begin{figure}[htp]
\includegraphics[width=0.8\linewidth]{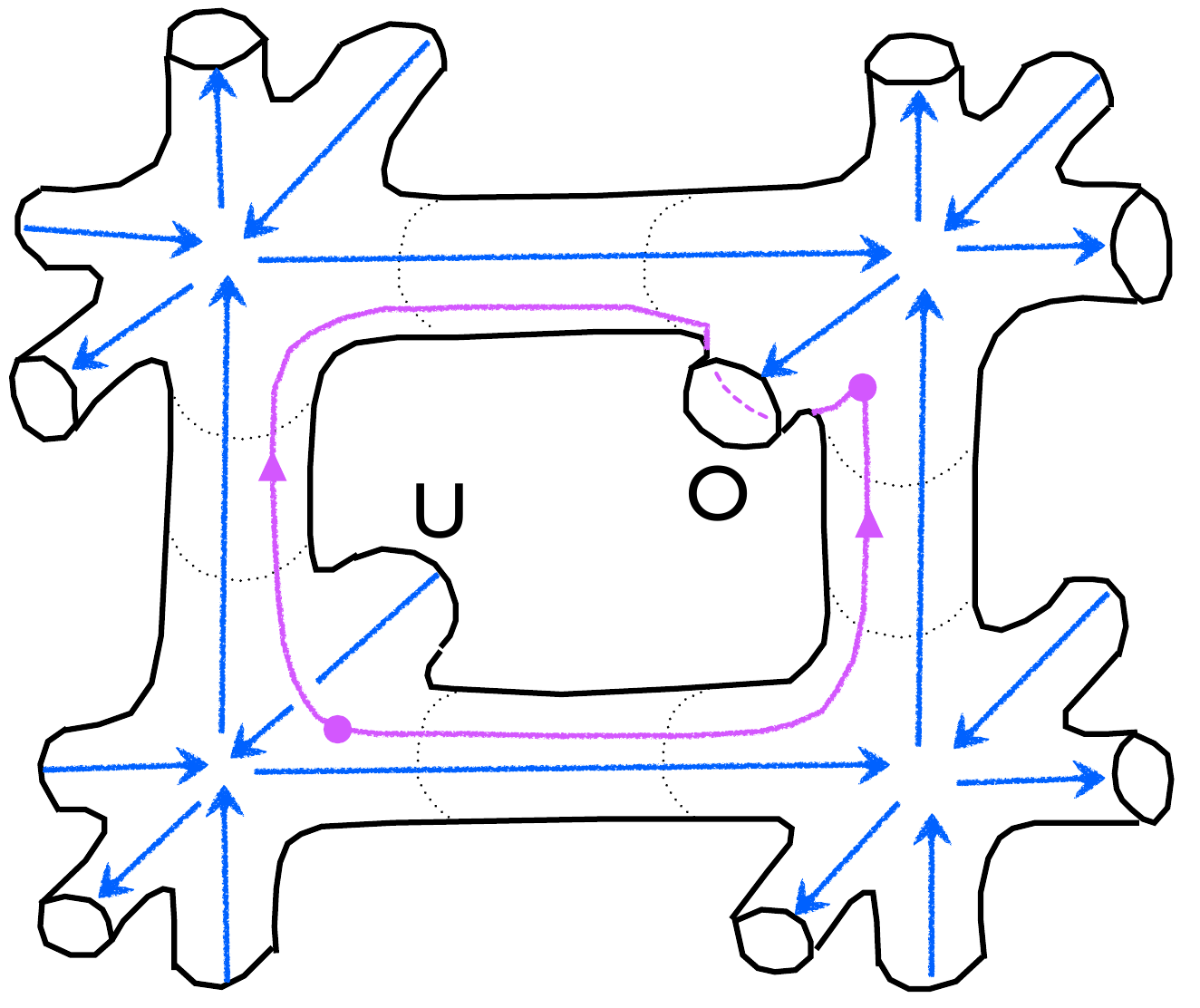}
 \caption{The blue arrows describe the nucleation, transport, and fusion and splitting process performed to construct a basis state in the 2D chiral theory with specified, well defined link quantum numbers.  The purple arrows describe the process corresponding to a plaquette term.  Expressing this plaquette term in the aforementioned basis amounts to fusing the purple and blue processes using associativity and braid phases, exactly as in \cite{Walker2012}.  These phases lead precisely to the extra signs associated with the $O$ and $U$ links in eqs. \ref{eq:bpe} and \ref{eq:bpm}.  
}
\label{thickened_p}
 \end{figure}

However, this identification of link labels with topological charges does not completely determine the identification of Hilbert spaces, because of a phase ambiguity: specifying the topological charge flowing through each link only determines a ground state of the 2D chiral theory up to an overall phase.  To fix this phase, we construct, for each choice of link labels, the corresponding state in the 2D chiral theory by starting with the trivial state (trivial topological charges through all links), and nucleating, transporting, and fusing and splitting anyons in the appropriate way through all the links.   Some arbitrary choices have to be made in this procedure: Specifically, we use the 2D projection introduced above, and take the process to proceed upward with respect to this projection, with fusion and splitting occurring on the top of the surface around each vertex, as in Fig. \ref{thickened_p}.  These choices reflect the gauge ambiguities inherent in the Walker Wang Hamiltonian.   (Boundary conditions consistent with those of the lattice Hamiltonian also need to be imposed, but this is straightforward.)

The key feature of this seemingly complicated construction is that the plaquette terms in (\ref{Ham}) take an extremely simple form in the 2D chiral theory: $B_P^{(a)}$ is just given by nucleating a pair of $a$ anyons, transporting one on the minimal girth path around the hole corresponding to $P$, and re-annihilating.  Indeed, to express this latter operator in the above basis amounts to fusing the purple path in figure \ref{thickened_p}, to the blue path, and this is accomplished with associativity and braiding phases (i.e. $F$ and $R$ moves), precisely as in the general definition of the Walker Wang model \cite{Walker2012}.  In this case, these associativity and braiding phases just amount to the extra signs associated with the $O$ and $U$ links.

The uniqueness of the ground state in this system now readily follows.  Indeed, we can equally well think of the plumber's nightmare surface $M$ as being associated with the dual lattice (Fig. \ref{pnightmare}); the original plaquette terms simply measure the flow of topological charge along the dual links, and imposing all of them just determines the unique state where all of these topological charges are $0$.  In other words, imposing trivial topological charge through these dual links is tatamount to cutting them and reducing $M$ to a product of spheres, a topology which hosts a unique ground state.

\subsection{ No deconfined bulk excitations}  
\begin{figure}[htp]
\includegraphics[width=0.8\linewidth]{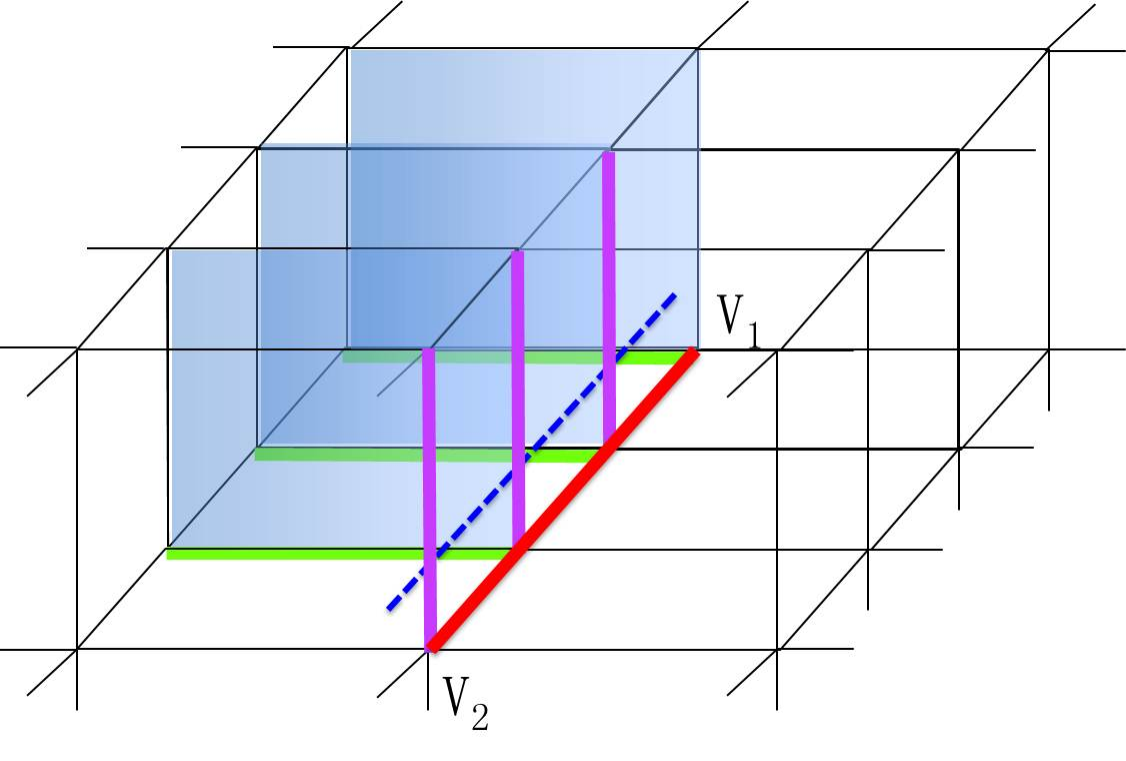}
 \caption{ Excitations in the bulk are confined.
 The path $C_{12}$ is shown in red; the displaced path used to determine $^*C_{12}$ is indicated with a dashed blue line.  Links that cross under (over) this path are colored green (purple).  The violated plaquettes (shaded blue) are those that are threaded by the dashed blue line.}
\label{ConfinedFig}
 \end{figure}

In the previous section we argued that the bulk of our model has no topological order; in particular, this means that there should be no non-trivial deconfined bulk excitations.  However, at least in two dimensions \cite{Levin2005}, string-net models possess string operators that commute with the Hamiltonian everywhere except at the endpoints, and hence create deconfined excitations.  To see how our 3D model evades this, consider the simplest way to create a pair of say $e$ charges at points $1$ and $2$, for simplicity separated only along the $y$ direction, namely acting with $\prod_{i\in C_{12}}\sigma^z_i$, where $C_{12}$ is a path of links connecting $1$ and $2$ (see Fig. \ref{ConfinedFig}).  We immediately see that this operator fails to commute with certain $xz$ plaquettes whose $O$ or $U$ links lie in $C_{12}$; the number of affected plaquettes is proportional to the length of $C_{12}$, leading to a linear confinement energy.

One can attempt to do better with the modified string operator
\be \label{oSV12}
{\bf S}^e_{12} = \prod_{i \in C_{12} } \sigma^z_i \prod_{j \in ^*C^{\rm over}_{12}} \sigma^x_j \prod_{k \in ^*C^{\rm under}_{12}}\sigma^x_k \tau^x_k
\ee
where $^*C_{12}$ is the set of links that are crossed by a curve that runs parallel to $C_{12}$, but is offset infinitesimally in the $-\hat{x} + \hat{y} + \hat{z}$ direction, and $^*C_{12}^{\rm over/under}$ are the subsets of links that cross over (under) the path $C_{12}$ in our projection (colored purple and green respectively in Fig. \ref{ConfinedFig}).  This new string operator fails to commute precisely with the blue shaded plaquettes in Fig. \ref{ConfinedFig}, and it is not possible to further reduce the number of violated plaquettes for a given $C_{12}$\cite{Walker2012,Keyserlingk2013}.  Hence bulk excitations carrying non-trivial gauge charge are linearly confined.

\begin{figure}[htp]
\includegraphics[width=0.8\linewidth]{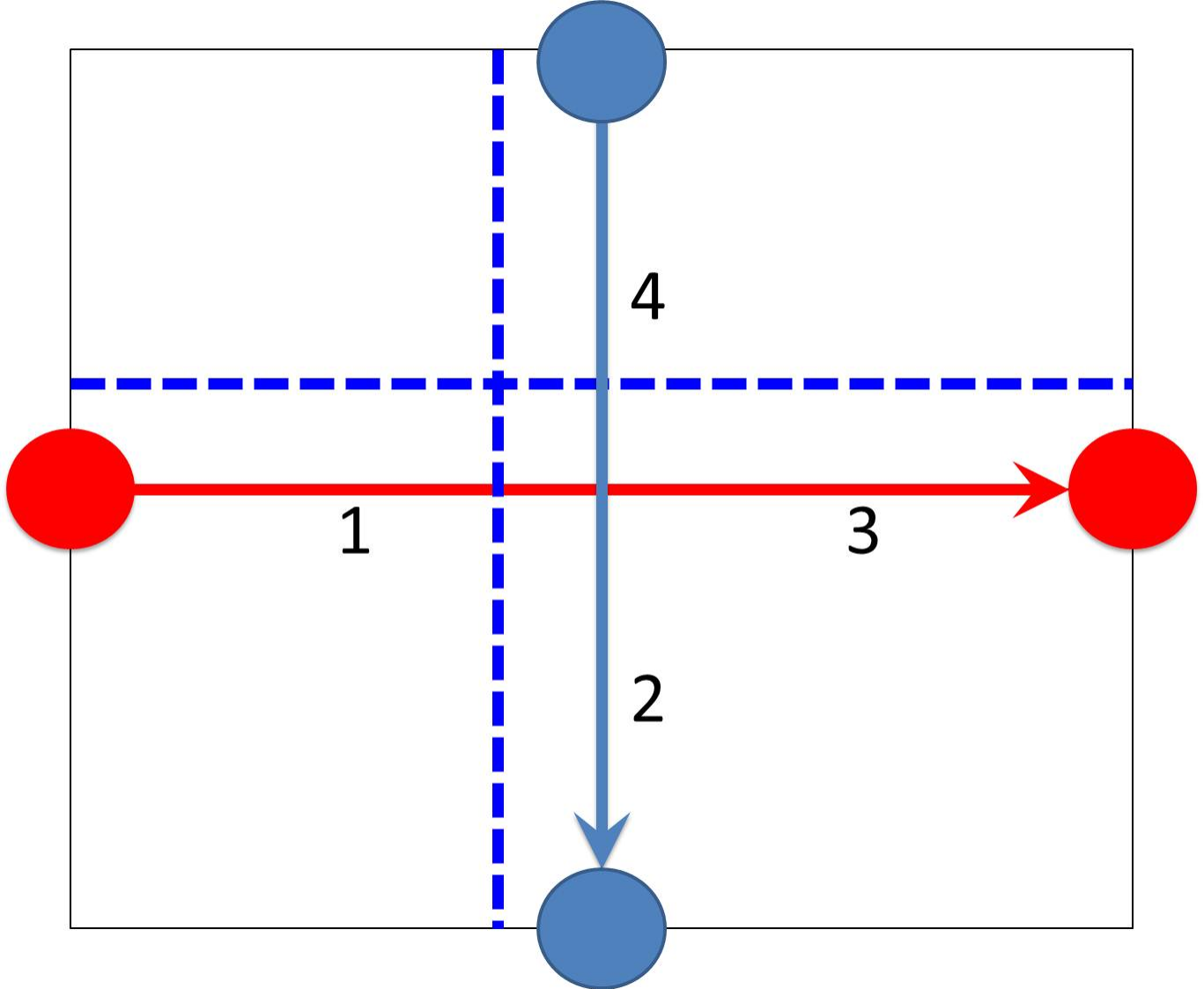}
 \caption{ Mutual Statistics: The operation of braiding a pair of anyons (say the $ e$ and $m$ particles) is captured by first creating a pair of $e$ particles (red) followed by a pair of $m$ particles (blue) in the manner shown. We now annihilate first the $e$ and then the $m$ particles to return to the vacuum, and examine the resulting phase. }
\label{MutualStatistics}
 \end{figure}

\subsection{ Deconfined Surface Excitations:}  Observe that if we terminate the system at the $xy$ plane of the curve $C_{12}$ in Fig. \ref{ConfinedFig}, the defective (blue shaded) plaquettes would not be included in the 3D lattice.  Indeed, retaining only the links below and including this $xy$ plane still gives an exactly soluble Hamiltonian, with surface vertex and plaquette terms involving only 5 links each, and now the string operator
\be \label{oSV12_surface}
{\bf S}^e_{surf.} = \prod_{i \in C_{12} } \sigma^z_i \prod_{k \in ^*C^{\rm under}_{12}}\sigma^x_k \tau^x_k
\ee
commutes with the Hamiltonian away from points $1$ and $2$, so that the $e$ charges it creates are deconfined.  Similarly
\begin{align}
{\bf S}^m_{surf.} &=& \prod_{i \in C_{12} } \tau^z_i \prod_{k \in ^*C^{\rm under}_{12}} \tau^x_k &\\
{\bf S}^\ve_{surf.} &=& \prod_{i \in C_{12} } \left(  \sigma^z_i \tau^z_i \right) \prod_{k \in ^*C^{\rm under}_{12}}\sigma^x_k &
\end{align}
create deconfined $m$ and $\ve$ excitations at the surface.

Let us examine the statistics of these excitations.  For definiteness, consider first the full braid of $e$ and $m$.  The statistical phase can be obtained by first nucleating a pair of $e$ particles, then a pair of $m$ particles as shown in Figure \ref{MutualStatistics}, and next annihilating first the $e$ particles and then the $m$ particles.  It is readily seen, using the explicit form of the surface string operators constructed above, that the product of the four corresponding string operators is $-I$, demonstrating the mutual semionic statistics of $e$ and $m$.  



\begin{figure}[htp]
\includegraphics[width=0.9\linewidth]{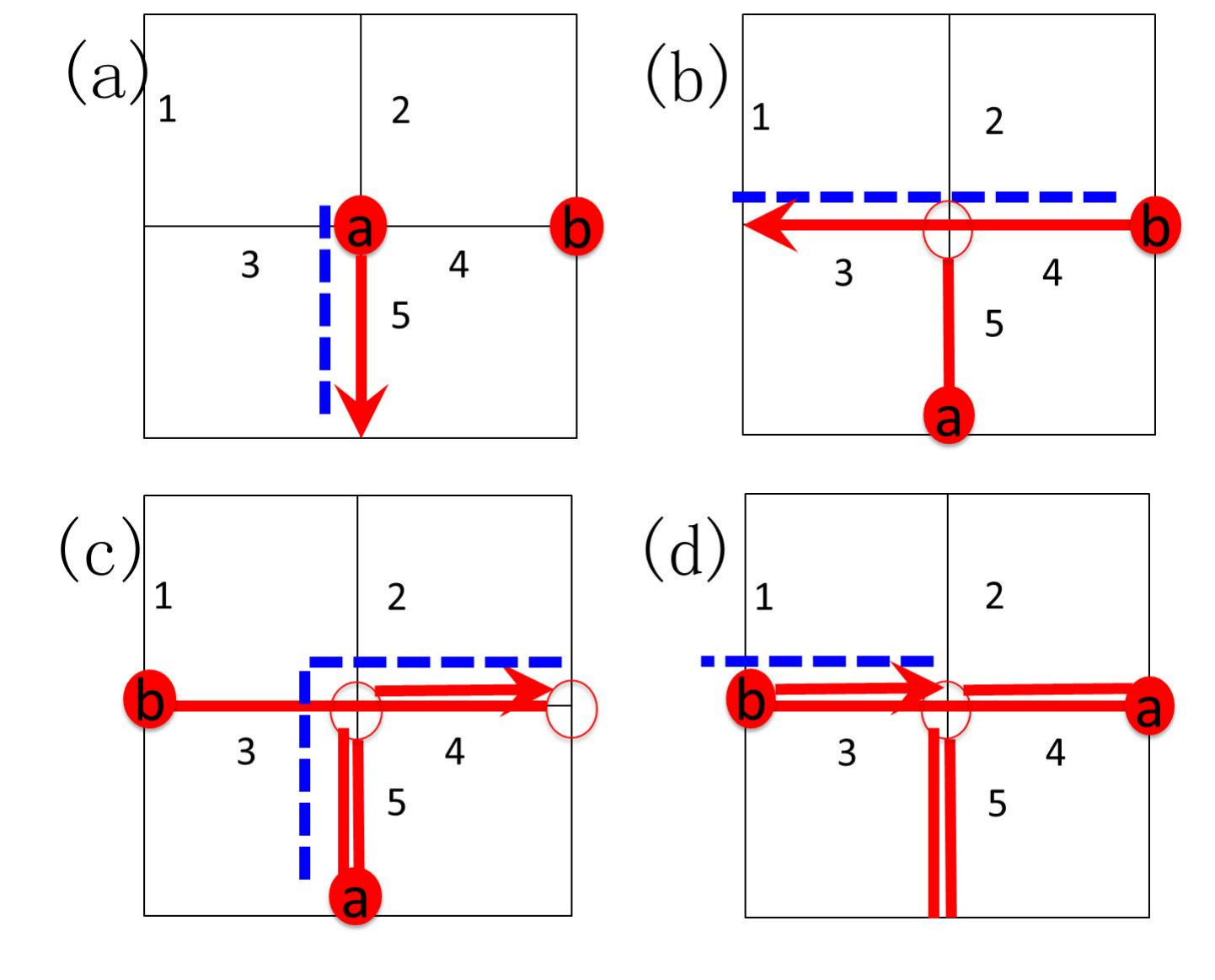}
 \caption{ Fermionic statistics:  The sequence of operations described in the text to exchange a pair of fermions.  The positions of the fermions are indicated by the red dots.  (The open red circles in (b) - (d) indicate the original positions of the fermions).  Solid red arrows indicate the link acted on by a string operator to move the fermion at this step; the solid red lines show where the string operators have acted at previous steps.  The links crossed by the dashed blue line are in $^*C$: there is a phase of $-1$ every time a dashed blue line crosses a solid red line.  
 }
\label{ExchangeFig}
 \end{figure}

 Next, consider exchanging two anyons of the same type, which can be carried out as shown in Fig.\ref{ExchangeFig}. We begin with two anyons (labelled $a$ and $b$) of the same type at vertices $i$ and $i+x$ respectively. 
The first step in the exchange is to move the anyon $a$ from $i$ to $i-y$; then move anyon $b$ from $i+x$ to $i-x$; next move anyon $a$ from $i-y$ to $i+x$; and finally move the anyon $b$ from $i-x$ to $i$. 
This process exchanges the two anyons. The whole procedure is realized by the string operator 
\be
 \oS^{\mu}_{ C_{i - \hat{x}, i}} \oS^{\mu}_{ C_{i - \hat{y}, i+ \hat{x}}}  \oS^{\mu}_{C_{i+ \hat{x}, i - \hat{x}}}  \oS^{\mu}_{C_{i, i-\hat{y} }} \ .
 \ee 
 We can explicitly check that this operator is equal to $-I$ for $\mu=e$,$m$,$\ve$. 
 For example, when $\mu=e$, $\oS^e_{C_{i, i-\hat{y} }}=\sigma^z_5\sigma^x_3\tau^x_3$, $\oS^e_{C_{i+ \hat{x}, i - \hat{x}}}=\sigma^z_3\sigma^z_4\sigma^x_1\tau^x_1\sigma^x_2\tau^x_2$, $\oS^e_{ C_{i - \hat{y}, i+ \hat{x}}}=\sigma^z_5\sigma^z_4\sigma^x_3\tau^x_3\sigma^x_2\tau^x_2$, $\oS^e_{ C_{i - \hat{x}, i}}=\sigma^z_3\sigma^x_1\tau^x_1$. 
 The total exchange string operator is then equal to $-I$. Similar checks can be performed for $\mu=m$ and $\ve$.

We note that the existence of the three fermion surface topological order can also be seen from the plumber's nightmare picture of our system: applying the bulk plaquette operators on a system with boundary leaves us with a bulk made out of disconnected spheres, and a surface with precisely the three fermion topological order.  However, the time reversal symmetry ${\mathcal{T}}$ cannot be easily understood in this picture.

\section{ Discussion:}

We have shown how to construct an exactly solvable lattice Hamiltonian that realizes the topologically ordered 3-fermion surface state in a time-reversal invariant way. It is important to emphasize that any purely 2+1D realization of this surface state necessarily breaks time reversal.  This follows from the relation
\be
\frac1{\mathcal D}\sum_a d_a^2\theta_a=e^{i2\pi c_-/8}
\ee
between anyons and chiral central charge, valid for any gapped 2D bosonic system, where ${\mathcal D} =\sqrt {\sum_a d_a^2}$. For the three fermion state where $d_a=1$ and $\theta_a= \{1,\,-1,\,-1,\,-1\}$ this requires $c_-= 4\, (\text{mod}\, 8)$, i.e. protected chiral edge modes.  Likewise, our 3+1D surface realization of this state is an indication of the non-trivial nature of the bulk SPT phase.  Indeed, we can destroy the topological order in a surface domain by adding a layer of a 2D (${\mathcal T}$-breaking) realization and condensing pairs; doing so in the opposite $\mathcal T$ breaking way on a bordering domain generates a $c_-=4-(-4)=8$ chiral mode, indicative of a 3+1D BTSc \cite{Vishwanath2012}.  Equivalently, the $\mathcal T$ broken surface displays a thermal analogue of the quantized magneto-electric effect \cite{Hasan2010,Hasan2011,Qi2011}.

More generally, realizing a topologically ordered phase which transforms under symmetry in a way that is forbidden in 2D necessarily leads to a protected surface state.  Let us illustrate this for the 3 fermion state by assuming the opposite is true-- i.e. the 3-fermion surface state can be eliminated without breaking $\mathcal T$ symmetry. Then, one can make a slab of the 3D phase with well separated top and bottom surfaces, and eliminate the surface state on the bottom side. Now, consider shrinking the slab until the 2D limit is reached. Since the bottom side and bulk are gapped, it should be possible to retain the original surface state on the top surface, without changing the symmetry. This produces a 2D realization of the `impossible' 2D state; therefore our assumption that it is possible to eliminate the surface state without breaking the symmetry must be false.  

Our construction is one member of a general class of 3+1D models constructed by Walker and Wang \cite{Walker2012,Keyserlingk2013}.  As we explain in detail in Appendix \ref{Appb}, their prescription allows one to turn a topologically ordered surface state - encoded in a unitary modular tensor category (UMTC) \cite{Note6}  - into a 3+1D bulk Hamiltonian and ground state wave function.  The latter is a superposition of loops (more precisely `string nets' \cite{Levin2005}) labeled by the anyons of the theory. The amplitude $\Psi_{\rm 3D}(C)$ for a given string net configuration $C$ is determined by the expectation value of the corresponding Wilson loop operators in the 2+1D TQFT; i.e.:
\be \label{eq:bulk_state}
\Psi_{\rm 3D}(C) = \langle W(C)\rangle_{\rm 2+1TQFT}
\ee
This is similar in spirit to e.g. Quantum Hall wave functions, which are related to the space-time correlations of their edge states. Here, since we demand a topologically ordered boundary state, the expectation values are taken in the boundary TQFT.  We emphasize that in general, these Walker-Wang models (built from a UMTC) have no bulk topological order, but, as stressed in this paper, the imposition of a symmetry can turn such a model into a non-trivial SPT.  Another example of such a construction is given in Ref. \cite{ProjectiveSemion} where an a chiral spin liquid with an anomalous realization of $\mathbb{Z}_2 \times \mathbb{Z}_2$ symmetry is realized as the surface of a decorated Walker-Wang model.

The string net picture also gives us intuition for the linear confinement of bulk quasiparticles.  Indeed, according to (\ref{eq:bulk_state}) a long string in the bulk will change the quantum fluctuation phase factors of small loops along its length by their relative braiding phase, when the small loop encircles the long string.  Because we have a UMTC, at least some of these braiding phases must be non-trivial, leading to a finite energy cost.  However, open strings lying on the surface, where no loops can encircle them, may give rise to deconfined excitations.

\begin{acknowledgments} We thank P. Dumitrescu, M.P.A. Fisher, A. Kitaev, Y-M. Lu, M. Metlitski, T. Senthil, and X-G. Wen, for helpful discussions.  In particular, we thank M.P.A. Fisher for introducing to us the idea of the ``plumber's nightmare" geometry.  A.V. is supported by ARO MURI Grant W911-NF-12-0461 and  X.C. by the Miller Institute for Basic Research in Science at Berkeley.  L.F. and F.B. are grateful for the hospitality of KITP (made possible by NSF Grant No. NSF PHY11-25915). Near the completion of this work we learnt of two preprints \cite{Metlitski2013} \cite{Wang2013} on 3D SPT phases. The former utilizes the statistical Witten effect to cleverly constrain the surface topological order while the latter uses an ingenious construction to obtain topologically ordered surface states for various SPT phases. 
\end{acknowledgments}

\begin{appendix}

\section{The 2D 3-fermion anyon model}  \label{AppA}


Our lattice Hamiltonian is related to a 2+1D anyon model with three types of fermions, which we will describe in more detail here.  
We may think of these three fermions as fermionic $\mathbb{Z}_2$ charges ($e$), fermionic $\mathbb{Z}_2$ fluxes ($m$), and a bound state of a charge and a flux ($\ve$).  Because the charges acquire a $\pi$ Berry phase upon encircling the $\mathbb{Z}_2$ fluxes, it can be checked that $\ve$ is also a fermion. Moreover, the three species of anyonic excitations all have mutual semionic statistics, i.e. braiding one around another induces a phase factor of $-1$.  Due to the symmetry in these statistics, we will also refer to this topological state the `three fermion $\mathbb{Z}_2$ model'; in practice it is irrelevant which one of the labels  
$\{e,m,\ve\}$ we assign to the flux and which to the `original' fermionic charge. We will use the label $1$ to designate the vacuum.

An explicit field theory of this 2D state can be written using an Abelian Chern Simons theory with four $U(1)$ gauge fields: \begin{eqnarray}
4\pi S_{\rm TQFT} &=& \int d^3 x \sum_{I,J=1}^4 K^{SO(8)}_{IJ} \epsilon^{\mu\nu\lambda} a^I_\mu \partial_\nu a^J_\lambda \\
K^{SO(8)} &=& \left (  \begin{array}{cccc} 2 &-1 &-1 & -1 \\ -1 & 2 & 0 & 0\\ -1 & 0 & 2 & 0 \\ -1 & 0& 0& 2 \end{array}\right )
\end{eqnarray}
This is the Cartan matrix of $SO(8)$. Note that the inverse matrix is:

\be
\left [{K^{SO(8)}}\right ]^{-1} = \left (  \begin{array}{cccc} 2 & 1 & 1 &  1 \\ 1 & 1 & \frac12 & \frac12\\ 1 & \frac12 & 1 & \frac12 \\ 1 & \frac12 & \frac12 & 1 \end{array}\right )
\ee
which clearly demonstrates the mutual statistics of the 3 fermion model, obtained by inner products $\theta_{ij} = 2\pi l_i^T\cdot K^{-1} \cdot l_j$, while self statistics is given by: $\theta_i = \pi l_i^T\cdot K^{-1} \cdot l_i$ where $l_i$ are integer vectors representing the quasiparticles. The eigenvalues of $K^{SO(8)}$ are all positive, indicating that all four edge modes are chiral (propagate in the same direction). Therefore, as emphasized in the main text, the 3 fermion state explicitly breaks time reversal symmetry, when realized in 2D.

\section{From 2D anyon model to lattice Hamiltonian - Technical Details and Physical Picture}  \label{Appb}
The lattice Hamiltonian presented in the main text is based on
a general construction introduced by Walker and Wang\cite{Walker2012,Keyserlingk2013}, applied to the specific case of the 3-fermion model discussed above.  Here we will give a qualitative description of the ground states that result from the Walker-Wang construction, and explain how it gives rise to the trivial bulk and topologically ordered surface of our lattice Hamiltonian.

A Walker-Wang model can be built from any anyon model.  The Hilbert space consists of all ways of assigning an anyon label to each edge of the lattice; the Hamiltonian is chosen such that the ground state wave function is a superposition of loops (more precisely ``string nets", in the sense of \cite{Levin2005}) labelled by the anyon types. The amplitude for a given configuration of these loops $C$ in the 3+1D wave function $\Psi_{\rm 3D}(C)$ is determined by the expectation value of the corresponding Wilson loop operators  in the 2+1D TQFT; i.e. we set:
$$
\Psi_{\rm 3D}(C) = \langle W(C)\rangle_{\rm 2+1TQFT}
$$
This is similar in spirit to eg. Quantum Hall wavefunctions, which are related to the space-time correlations of their edge states. Here, since we demand a topologically ordered boundary state, the expectation values are taken in the boundary TQFT. Below we will argue more physically why the bulk wave function encodes the statistical interactions of the surface anyons, while posessing no topological order itself. 


\begin{figure}[htbp]\vspace{-0pt}
\includegraphics[width=0.4\linewidth]{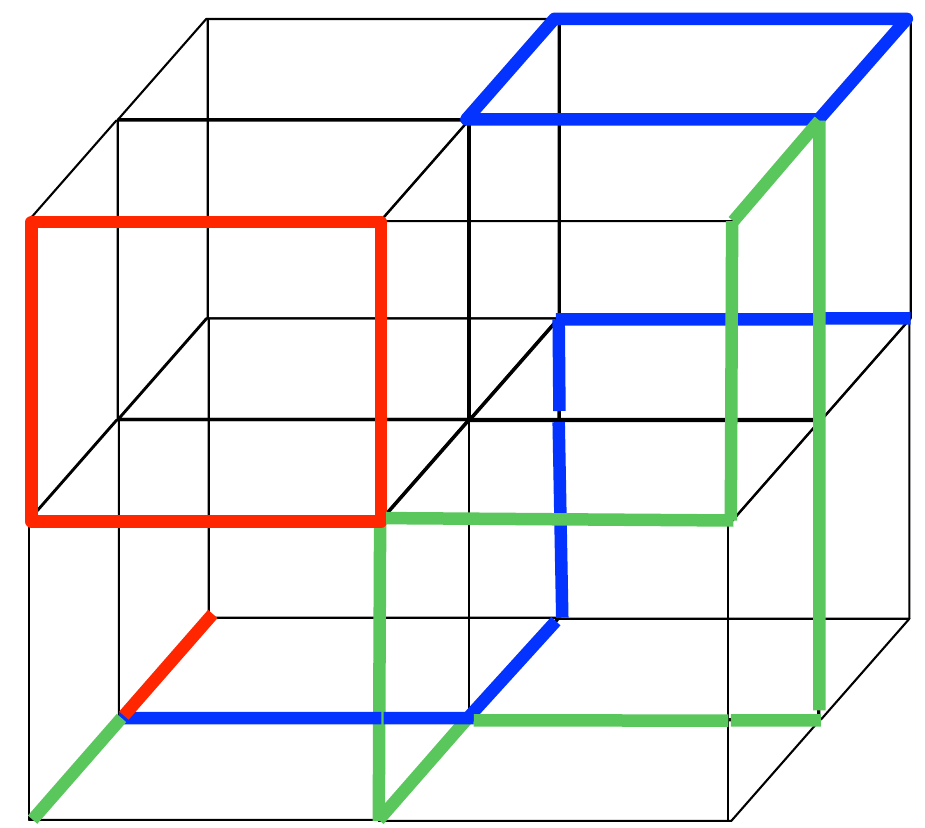} \ \ \ \ \ 
\includegraphics[width=0.4\linewidth]{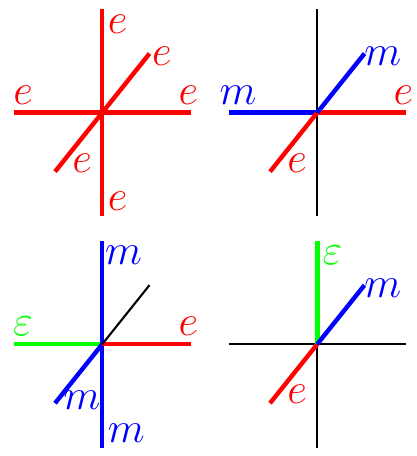}
\caption{\label{schematic} (color online) The Low energy Hilbert space of the lattice model consists of loops of three colors that satisfy fusion rules at the vertices - i.e. they are either closed loops of a single color, or segments of three colors can meet at a vertex. }
\end{figure}

The ground state wave function for the `3-fermion' Walker-Wang model hence contains three different colors of loops corresponding to the three species of fermions, as shown in Fig.\ref{schematic}. 
Any two colors can merge into the third, in accordance with the fusion rules of the TQFT:
\be
e\times e = 1 \ \ \ m \times m =1 \ \ \ \ve \times \ve = 1 \ \ \
e\times m = \ve 
\ee
These equations simply reiterate the fact that $e$ is a $\mathbb{Z}_2$ charge, $m$ is a $\mathbb{Z}_2$ flux, and $\ve$ is the combination of these two.

\begin{figure}[htbp]\vspace{-0pt}
\includegraphics[width=0.8\linewidth]{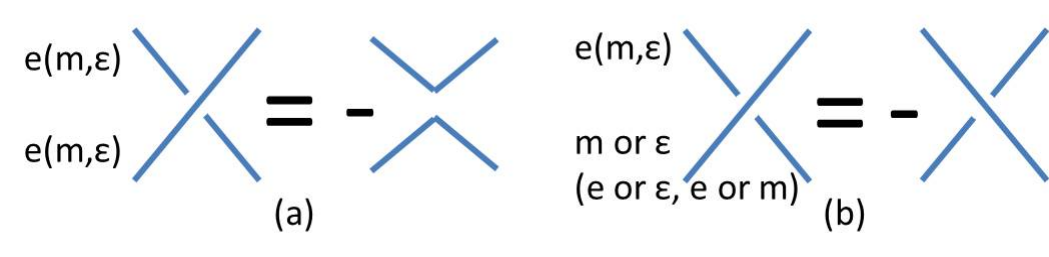}
\caption{(color online) Braiding rules for strings in the ground state wave funciton. (a) applies to strings of the same color while (b) applies to strings of different colors.}
\label{braiding}
\end{figure}

 Each loop configuration comes with a specific phase due to the twisting and intertwining of the fermion world-lines.  Since this twisting may depend on the angle of view, in order to calculate the phase, we need to pick a particular projection of the 3D wolrd-lines onto 2D ones.  The projection we will use is shown in Fig.\ref{schematic}.  Having fixed a projection, the phase factor can be obtained using the braiding rules (as shown in Fig.\ref{braiding}) given by the $R$ matrix:\cite{Note4}
\be
\begin{array}{l} \label{Req}
R_{\mu,\mu}=-1,\ \ \mu=e,m,\ve \\
R_{e,m}=R_{m,\ve}=R_{\ve,e}=-1 \\
R_{m,e}=R_{\ve,m}=R_{e,\ve}=1
\end{array}
\ee
Isolated loops can shrink to the vacuum without an extra phase factor (the quantum dimension of our fermions is 1). The ground state wave function is a superposition of all allowed loop configurations weighted by the corresponding phase factor. When the system has a surface, the same graphical rules can be used to determine the wave function for the loop gas after fixing the projection.

An important feature of this state is that, because the braiding rules involve no complex numbers, the ground state wave function is {\em real}, hence symmetric under the time reversal operator $\mathcal T$ that acts by complex conjugation. This is even true for the wave function on a 3D manifold with boundary. Therefore, no time reversal symmetry breaking occurs either in the bulk or on the surface.

\subsection{Understanding bulk and surface theories from the wave-function}

We can also gain an intuitive understanding of the bulk confinement and surface deconfinement of our lattice model by considering the relationship between the low-energy states of the Walker-Wang model and anyon world-lines in the 3-fermion model.  We note in passing that the Hamiltonian given in the main text differs slightly from the Walker-Wang construction, which has a third component to the plaquette term.  However, this only affects the relative energies of the excitations, and is not important for any of the qualitative features discussed here.

\begin{figure}[htbp]\vspace{-0pt}
\includegraphics[width=0.6\linewidth]{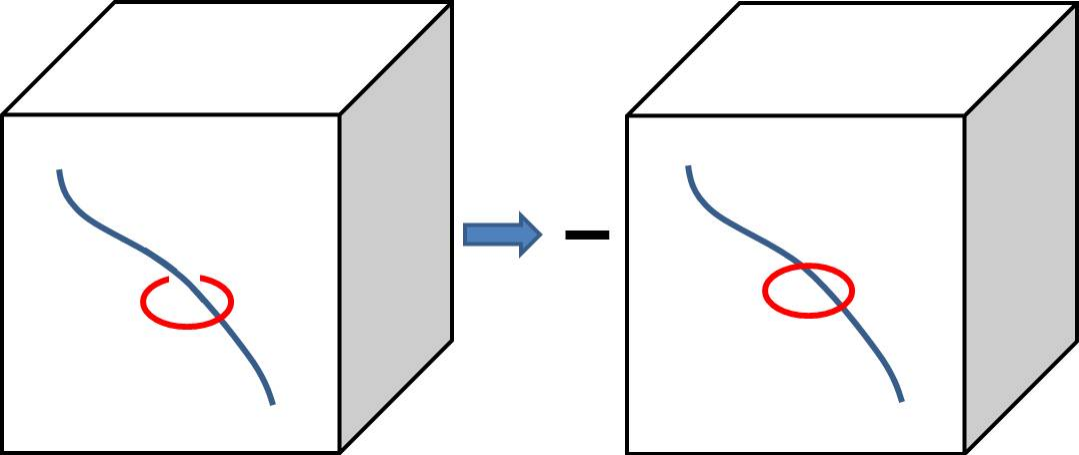}
\caption{(color online) Open strings in the bulk change the quantum fluctuation phase factors of loops along its length, which costs finite energy. Therefore, the end of strings are confined in the bulk.}
\label{confinement}
\end{figure}

 Let us begin with the bulk.  We can create excitations by adding open
 strings  to the ground-state string net.  In the bulk, however, the excitation energy grows linearly with the string length, leading to confinement of the particles at the ends of the strings.
To see the confinement, consider an open string (for example blue) in the bulk which is circled by a small ring of a different color (for example red), as shown in Fig.\ref{confinement}
The braiding rules dictate that un-linking this ring from the open string will result in a $(-)$ sign in the wave-function.  That is, introducing the open string changes the phase factors associated with small fluctuating loops along its length, which costs finite energy. Therefore, the string's endpoints cannot be separated very far, and the fermionic excitations in the bulk are confined. 

More generally, any set of strings with consistent braiding and fusion rules given by a unitary braided fusion category can be used to write 3D string-net wave functions in a similar fashion. As shown in Ref.\cite{Keyserlingk2013}, the bulk of the state has no deconfined excitations, and hence no nontrivial topological order, as long as each string has nontrivial statistics with at least one of the other strings. The corresponding category is said to be `modular'. 

\begin{figure}[htbp]\vspace{-0pt}
\includegraphics[width=0.6\linewidth]{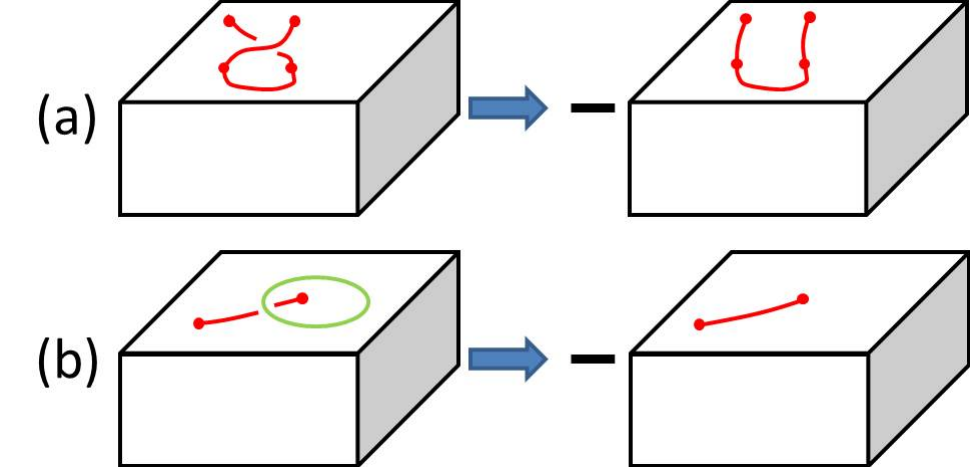}
\caption{(color online) The anyonic excitations on the surface are created by open strings. At the ends of the strings are three species of fermions (corresponding to three colors of open strings) which have mutual semionic statistics.  This can be seen from the braiding statistics of the strings generating (a) the exchange and (b) the braiding of the end of strings.}
\label{surface_anyon}
\end{figure}

However, the above argument suggests that open strings lying on the surface, where no loops can encircle them, may give rise to deconfined excitations, as we have verified explicitly for the 3-fermion model in the main text.  The excitations at the end of these open strings have anyonic statistics.  
To understand this, note that in the presence of an open string he wave function becomes a superposition of all string configurations in which strings  end at the positions of the excitations. We can therefore determine the statistics of the excitations by tracking these open strings. Suppose we exchange two string ends of the same color (say red, as show in Fig.\ref{surface_anyon} (a))  by crossing two red string segments to the surface.  (Fig.\ref{surface_anyon}(a) shows one possible configuration.)  This twist in the string configuration (relative to the string configuration before exchange) can be removed to bring the strings back to their original form, but this results in a $(-)$ sign. Therefore, exchanging end of strings of the same color adds a $(-)$ sign to the total wave function, which is equivalent to saying that the ends of the strings are fermions. Similarly one can check that string ends of different colors have mutual semionic statistics by braiding them with linked loops on the surface as shown in Fig.\ref{surface_anyon}.

\end{appendix}


\end{document}